\documentclass[12pt]{iopart}

\usepackage{iopams}  
\usepackage{cite}
\usepackage{cleveref}

\newtheorem{proposition}{Proposition}

\begin{document}

\title[On integrability aspects of the SUSY sine-Gordon equation]{On integrability aspects of the supersymmetric sine-Gordon equation}

\author{S Bertrand}

\address{Department of Mathematics and Statistics, Universit\'e de Montr\'eal,\\ Montr\'eal CP 6128 Succ. Centre-Ville (QC) H3C 3J7, Canada}
\ead{bertrans@crm.umontreal.ca}
\vspace{10pt}

\begin{abstract}
In this paper we study certain integrability properties of the supersymmetric sine-Gordon equation. We construct Lax pairs with their zero-curvature representations which are equivalent to the supersymmetric sine-Gordon equation. From the fermionic linear spectral problem, we derive coupled sets of super Riccati equations and the auto-B\"acklund transformation of the supersymmetric sine-Gordon equation. In addition, a detailed description of the associated Darboux transformation is presented and non-trivial super multisoliton solutions are constructed. These integrability properties allow us to provide new explicit geometric characterizations of the bosonic supersymmetric version of the Sym-Tafel formula for the immersion of surfaces in a Lie superalgebra. These characterizations are expressed only in terms of the independent bosonic and fermionic variables.
\end{abstract}


\vspace{2pc}
\noindent{\it Keywords}: Supersymmetric extension of the sine-Gordon equation, Supersymmetric versions of the B\"acklund and Darboux transformations, Solitonic surfaces, Supersymmetric version of the Sym-Tafel immersion formula.


\maketitle

\section{Introduction}\label{Intro}

Over the past four decades, supersymmetric (SUSY) integrable models have generated a great deal of interest in the literature of mathematical physics (see e.g. \cite{BJ01,CR83,HU06,JP00,LM91,Mathieu88,TL09,Popowicz97,CIN02,BLR89,AB90,BL90} and references therein). Their special properties, such as the existence of a linear spectral problem (LSP), the B\"acklund and Darboux transformations and infinite sets of conserved currents, have allowed the construction of analytical supersoliton solutions (e.g. \cite{GRC01,GHS09,LM98,SHS06,LM07,VF77,BLR89}). Some of these integrability properties have been investigated for the case of the SUSY sine-Gordon equation, see e.g. \cite{CK78,GYZ09,SH05,SHS06,LM98,LM07,FGS78,GS78,Sciuto80,BLR89}. The approach proposed in this paper goes deeper into certain integrability properties of the SUSY sine-Gordon equation than \cite{SH05,SHS06}, especially the Darboux transformation of the SUSY sine-Gordon equation, which will later enable us to obtain new results for the bosonic SUSY version of the Sym-Tafel formula for the immersion of surfaces in Lie superalgebras. One should note that $n$-order Darboux transformations were investigated in \cite{LM98,LM07} using Pfaffian solutions. The associated linear system uses $2\times2$ matrices which are more compact than our $3\times3$ potential supermatrices. However, the linear system used in these articles is not convenient for the bosonic SUSY version of the Sym--Tafel immersion formula since this system uses differential matrix operators instead of supermatrices, as used in our approach. In this paper, we discuss the links between different integrability properties associated with the SUSY sine-Gordon equation. Furthermore, we provide explicit solutions for the SUSY sine-Gordon equation and its LSP which allow us to investigate examples of the bosonic SUSY version of the Sym-Tafel immersion formula through their geometric characterization.

This paper is organized as follows. In section 2, we present some of the integrability properties of the SUSY sine-Gordon equation. We derive a fermionic LSP and we link it to different integrability properties, such as a bosonic version of the LSP, equivalent coupled sets of super Riccati equations and the auto-B\"acklund transformations of the SUSY sine-Gordon equation. Moreover, we provide a detailed description of the Darboux transformations associated with the SUSY sine-Gordon equation. In section 3, we investigate two examples of the bosonic SUSY version of the Sym-Tafel formula for immersion associated with the SUSY sine-Gordon equation. These examples are obtained using the first iteration of the Darboux transformation and are exclusively written in terms of the bosonic and fermionic independent variables. A characterization of the geometry of each surface is provided.

\section{Integrability aspects of the supersymmetric sine-Gordon equation}\label{SecAspect}
Throughout this paper, we follow the notation introduced in Section 3 of the paper \cite{BG16}. A more detailed presentation of the theory of Grassmann algebras can be found in the books \cite{Cornwell,DeWitt,Freed,Varadarajan} and references therein. In what follows we do not use the implicit notation for the fermionic derivatives of a ($m\vert n$)-supermatrix $M$ \cite{Cornwell}, e.g.
\begin{equation*}
\partial_{\theta}M=\left(\begin{array}{cc}
\partial_\theta A & \partial_\theta B \\
-\partial_\theta C & -\partial_\theta D
\end{array}\right).
\end{equation*}
Instead, we introduce a matrix $E$ such that
\begin{equation*}
E\partial_\theta M=\left(\begin{array}{cc}
\partial_\theta A & \partial_\theta B \\
-\partial_\theta C & -\partial_\theta D
\end{array}\right),\qquad E=\left(\begin{array}{cc}
I_m & 0 \\
0 & -I_n \\
\end{array}\right),
\end{equation*}
where the square submatrices $I_m$ and $I_n$ represent the identity matrices of dimension $m$ and $n$, respectively. The chain rule is assumed to be
\begin{equation*}
\frac{d}{dv}f(u)=\frac{df}{du}\frac{du}{dv}.
\end{equation*}

According to \cite{Coleman75}, the SUSY sine-Gordon equation (SSGE),
\begin{equation}
D_+D_-s=i\sin s,\label{SSGE}
\end{equation}
is considered for a bosonic superfield $s=s(\theta^+,\theta^-,x_+,x_-)$ with the covariant derivatives
\begin{equation}
D_\pm=\frac{\partial}{\partial \theta^\pm}-i\theta^\pm\frac{\partial}{\partial x_\pm},
\end{equation}
where $\theta^\pm$ are fermonic independent variables and $x_\pm$ are bosonic light-cone coordinates. The fermionic derivatives $D_\pm$ have the properties
\begin{equation}
D_\pm^2=-i\partial_{x_\pm},\qquad \lbrace D_+,D_-\rbrace=0,\label{Dprop}
\end{equation}
where $\lbrace\cdot,\cdot\rbrace$ stands for the anticommutator. The SSGE (\ref{SSGE}) can be obtained through the super Euler-Lagrange equation,
\begin{equation}
\frac{\partial}{\partial s}\mathcal{L}+D_+\left(\frac{\partial}{\partial (D_+s)}\mathcal{L}\right)+D_-\left(\frac{\partial}{\partial (D_-s)}\mathcal{L}\right)=0,
\end{equation}
with the Lagrangian density
\begin{equation}
\mathcal{L}=\cos s-\frac{i}{2}D_+sD_-s.
\end{equation}

The SSGE (\ref{SSGE}) is known to be integrable in the sense of soliton theory \cite{BLR89,CK78,SH05,SHS06,GS78,FGS78}. One can provide an infinite set of locally conserved quantities and a LSP under the form of a differential linear matrix representation. One way to obtain a linearization of the SSGE (\ref{SSGE}) is to consider the following problem for the wavefunction $\Phi$:
\begin{equation}
D_+\Phi=(\mathcal{J}e^{is}+\mathcal{K}e^{-is})\Phi,\qquad D_-\Phi=(\mathcal{M} D_-s+\mathcal{N})\Phi,\label{preLSP}
\end{equation}
where $\mathcal{J,K,M,N}$ are complex-valued matrices and then take the compatibility conditions of $\Phi$ to be equivalent to the SSGE (\ref{SSGE}). The resulting algebraic constraints are
\begin{equation}
\hspace{-2cm}i\mathcal{J}=[\mathcal{M},\mathcal{J}],\qquad i\mathcal{K}=[\mathcal{K},\mathcal{M}],\qquad \lbrace \mathcal{J},\mathcal{N}\rbrace=-\lbrace \mathcal{K},\mathcal{N}\rbrace,\qquad \frac{1}{2}\mathcal{M}=\lbrace \mathcal{K},\mathcal{N}\rbrace.
\end{equation}
One solution to these constraints, which takes its values in the $\mathfrak{sl}(3,\mathbb{C})$ Lie algebra, is
\begin{equation}
\begin{array}{ll}
\mathcal{J}=\frac{1}{2}\left(\begin{array}{ccc}
0 & 0 & i \\
0 & 0 & 0 \\
0 & 1 & 0
\end{array}\right),& \mathcal{K}=\frac{1}{2}\left(\begin{array}{ccc}
0 & 0 & 0 \\
0 & 0 & -i \\
-1 & 0 & 0
\end{array}\right),\\
\\
\mathcal{M}=\left(\begin{array}{ccc}
i & 0 & 0 \\
0 & -i & 0 \\
0 & 0 & 0
\end{array}\right),& \mathcal{N}=\left(\begin{array}{ccc}
0 & 0 & -i \\
0 & 0 & i \\
-1 & 1 & 0
\end{array}\right).
\end{array}
\end{equation}
To introduce the spectral parameter $\lambda$ in equations (\ref{preLSP}), as proposed in\cite{BGH152}, we apply the Lie point symmetry transformation of the SSGE (\ref{SSGE}),
\begin{equation}
\hspace{-2.5cm}\tilde{x}_+=\lambda x_+,\qquad\tilde{x}_-=\lambda^{-1}x_-,\qquad\tilde{\theta}^+=\lambda^{1/2}\theta^+,\qquad\tilde{\theta}^-=\lambda^{-1/2}\theta^-,\qquad \lambda=\pm e^{\mu},
\end{equation}
to the linear system, where $\mu$ is any bosonic constant in the Grassmann algebra. One should note that this scaling transformation does not leave the linear system (\ref{preLSP}) invariant. Hence, the LSP of the SSGE (\ref{SSGE}) takes the form
\begin{equation}
D_\pm\Phi(\lambda,s)=U_\pm(\lambda,s)\Phi(\lambda,s)\label{LSP}
\end{equation}
where $\lambda$ is the bosonic spectral parameter and $U_\pm$ are fermionic supermatrices taking values in the $\mathfrak{sl}(2\vert1,\mathbb{G})$ superalgebra, given by \cite{SHS06}
\begin{equation}
\hspace{-2cm}U_+=\frac{1}{2\sqrt{\lambda}}\left(\begin{array}{ccc}
0 & 0 & ie^{is} \\
0 & 0 & -ie^{-is} \\
-e^{-is} & e^{is} & 0
\end{array}\right),\qquad U_-=\left(\begin{array}{ccc}
iD_-s & 0 & -i\sqrt{\lambda} \\
0 & -iD_-s & i\sqrt{\lambda} \\
-\sqrt{\lambda} & \sqrt{\lambda} & 0
\end{array}\right).
\end{equation}
The wavefunction $\Phi$ is a ($2\vert1$)-supervector
\begin{equation}
\Phi=\left(\begin{array}{c}
\psi\\
\phi\\
\chi
\end{array}\right),\qquad (-1)^{\deg(\psi)}=(-1)^{\deg(\phi)}=(-1)^{\deg(\chi)+1},\label{ppp}
\end{equation}
such that either $\psi,\phi$ are bosonic superfields and $\chi$ is a fermionic superfield, or $\psi,\phi$ are fermionic superfields and $\chi$ is a bosonic superfield.

The LSP can also be defined using an invertible wavefunction $\Psi$ in the $GL(2\vert1,\mathbb{G})$ supergroup, 
\begin{equation}
D_\pm\Psi(\lambda,s)=U_\pm(\lambda,s)\Psi(\lambda,s).\label{LSP2}
\end{equation}
The compatibility condition of both LSPs (\ref{LSP}) and (\ref{LSP2}), i.e.
\begin{equation}
D_+U_-+D_-U_+-\lbrace EU_+,EU_-\rbrace=0,\qquad E=\left(\begin{array}{ccc}
1 & 0 & 0\\
0 & 1 & 0\\
0 & 0 & -1
\end{array}\right),\label{ZCC}
\end{equation}
are equivalent to the SSGE (\ref{SSGE}).

Moreover, a LSP can be described using the light-cone coordinate derivatives $\partial_{x_\pm}$. From the property (\ref{Dprop}), we write the bosonic version of the LSP exclusively in terms of the wavefunction $\Psi$ and the potential matrices $U_\pm$ from the fermionic version of the LSP, i.e.
\begin{equation}
\partial_{x_\pm}\Psi=i(D_\pm U_\pm-(EU_\pm)^2)\Psi=V_\pm\Psi,\label{LSPB}
\end{equation}
where the bosonic supermatrices $V_\pm$ take the forms
\begin{equation}
\begin{array}{l}
V_+=\frac{1}{2}\left(\begin{array}{ccc}
\frac{1}{2\lambda} & \frac{-1}{2\lambda}e^{2is} & \frac{-i}{\sqrt{\lambda}}e^{is}D_+s \\
\frac{-1}{2\lambda}e^{-2is} & \frac{1}{2\lambda} & \frac{-i}{\sqrt{\lambda}}e^{-is}D_+s \\
\frac{-1}{\sqrt{\lambda}}e^{-is}D_+s & \frac{-1}{\sqrt{\lambda}}e^{is}D_+s & \frac{1}{\lambda}
\end{array}\right)\in\mathfrak{sl}(2\vert1,\mathbb{G}),\\
\\
V_-=\left(\begin{array}{ccc}
i\partial_{x_-}s-\lambda & \lambda & -i\sqrt{\lambda}D_-s \\
\lambda & -i\partial_{x_-}s-\lambda & -i\sqrt{\lambda}D_-s \\
\sqrt{\lambda}D_-s & \sqrt{\lambda}D_-s & -2\lambda
\end{array}\right)\in\mathfrak{sl}(2\vert1,\mathbb{G}).
\end{array}
\end{equation}
The compatibility conditions of the LSP (\ref{LSPB}) correspond to the ``classical'' version of the zero-curvature conditions, i.e.
\begin{equation}
\partial_{x_+}V_--\partial_{x_-}V_++[V_-,V_+]=0,
\end{equation}
which is satisfied whenever the SSGE (\ref{SSGE}) is satisfied.

To obtain coupled sets of super Riccati equations \cite{SH05,FGS78}, we consider the quantities
\begin{equation}
p=\frac{\phi}{\psi},\qquad q=\frac{\chi}{\psi},
\end{equation}
where we take $\psi,\phi$ to be bosonic superfields and $\chi$ to be a fermionic superfield. By differentiating them, we get
\begin{equation}
\begin{array}{l}
D_+p=\frac{-i}{2\sqrt{\lambda}}e^{-is}q-\frac{i}{2\sqrt{\lambda}}e^{is}pq,\\
D_-p=-2i(D_-s)p+i\sqrt{\lambda}(1+p)q,
\end{array}
\end{equation}
together with
\begin{equation}
\begin{array}{l}
D_+q=\frac{-1}{2\sqrt{\lambda}}e^{-is}+\frac{1}{2\sqrt{\lambda}}e^{is}p,\\
D_-q=\sqrt{\lambda}(p-1)-i(D_-s)q,
\end{array}
\end{equation}
from which one can obtain an infinite set of locally conserved currents \cite{FGS78}. The compatibility conditions of both sets of equations are satisfied whenever $s$ is a solution of the SSGE (\ref{SSGE}).
Moreover, by setting 
\begin{equation}
p\rightarrow e^{-i(s+\tilde{s})},\qquad q\rightarrow fe^{-\frac{i}{2}(s+\tilde{s})},
\end{equation}
where $f$ is a fermionic superfield, we obtain the auto-B\"acklund transformations of the SSGE \cite{GYZ09,SH05}
\begin{equation}
\begin{array}{l}
D_+(s+\tilde{s})=\frac{f}{\sqrt{\lambda}}\cos\left(\frac{s-\tilde{s}}{2}\right),\\
D_-(s-\tilde{s})=2\sqrt{\lambda}f\cos\left(\frac{s+\tilde{s}}{2}\right),\\
D_+f=\frac{i}{\sqrt{\lambda}}\sin\left(\frac{s-\tilde{s}}{2}\right),\\
D_-f=-2i\sqrt{\lambda}\sin\left(\frac{s+\tilde{s}}{2}\right).
\end{array}\label{Backlund}
\end{equation}
The compatibility conditions of these equations are satisfied whenever $s$ and $\tilde{s}$ are solutions of the SSGE (\ref{SSGE}).
One should note that the auto-B\"acklund transformation (\ref{Backlund}) of the SSGE requires an additional fermionic function $f$ due to the oddness of the derivatives $D_\pm$.

It is possible to construct $n$ soliton solutions of the SSGE (\ref{SSGE}) from the Darboux transformation using one (trivial) solution of the SSGE together with $n$ solutions of the associated LSP (\ref{LSP}) for $n$ fixed spectral parameters $\lambda_j$, $j=0,1,...,n-1$. One should note that Darboux transformations do not ensure that the newly constructed solutions are linearly independent of the previously constructed solutions. The first iteration of the Darboux transformation for the SSGE (\ref{SSGE}) \cite{SHS06,LM07,LM98} (similarly to the classical case \cite{AY95,MS91}) is given by
\begin{equation}
s[1]=s-i\ln\left(\frac{\psi_0}{\phi_0}\right),\label{sol1}
\end{equation}
\begin{equation}
\hspace{-2cm}
\Phi_j[1]=\left(\begin{array}{c}
\psi_j[1]\\
\phi_j[1]\\
\chi_j[1]
\end{array}\right)=\left(\begin{array}{ccc}
-\lambda_0\frac{\phi_0}{\psi_0} & \lambda_j & -i\sqrt{\lambda_0\lambda_j}\frac{\chi_0}{\psi_0} \\
\lambda_j & -\lambda_0\frac{\psi_0}{\phi_0} & -i\sqrt{\lambda_0\lambda_j}\frac{\chi_0}{\phi_0} \\
\sqrt{\lambda_0\lambda_j}\frac{\chi_0}{\psi_0} & \sqrt{\lambda_0\lambda_j}\frac{\chi_0}{\phi_0} & -(\lambda_0+\lambda_j)
\end{array}\right)\left(\begin{array}{c}
\psi_j\\
\phi_j\\
\chi_j
\end{array}\right),\label{Dar1}
\end{equation}
where $s$ is a solution of the SSGE (\ref{SSGE}), $\psi_j,\phi_j$ are bosonic solutions and $\chi_j$ is a fermionic solution of the LSP (\ref{LSP}) for the solution $s$ and the fixed spectral parameter $\lambda=\lambda_j$. The new solution $\Phi_j[1]$ of the LSP is given for the system
\begin{equation}
D_\pm\Phi_j[1]=A_\pm(\lambda_j,s[1])\Phi_j[1].\qquad j=1,2,3,...\label{LSP3}
\end{equation}
One should note that the solution of the LSP (\ref{LSP3}) with the fixed parameter $\lambda_0$ has been used in order to construct the new solution. The index $j=0$ for the first (or higher) iteration transformation correspond to the trivial solution $\Phi=0$. Therefore, the solution for the LSP with $\lambda=\lambda_0$ cannot be used to obtain other new solutions.

\begin{figure}
\setlength{\unitlength}{1mm}
\centering
\begin{picture}(100,120)
\put(11,105){$\Phi_n$}
\put(23,106){$\dots$}
\put(37,105){$\Phi_0$}
\put(65,105){$s$}
\put(14,103){\vector(0,-2){8}}
\put(36,103){\vector(-3,-2){14}}
\put(44,103){\vector(3,-2){14}}
\put(66,103){\vector(0,-2){8}}
\put(9,90){$\Phi_n[1]$}
\put(23,91){$\dots$}
\put(35,90){$\Phi_1[1]$}
\put(63,90){$s[1]$}
\put(14,88){\vector(0,-2){8}}
\put(36,88){\vector(-3,-2){14}}
\put(44,88){\vector(3,-2){14}}
\put(66,88){\vector(0,-2){8}}
\put(9,75){$\Phi_n[2]$}
\put(23,76){$\dots$}
\put(35,75){$\Phi_2[2]$}
\put(63,75){$s[2]$}
\put(13,64){$\vdots$}
\put(39,64){$\vdots$}
\put(66,64){$\vdots$}
\put(3,55){$\Phi_n[n-2]$}
\put(23,56){$\dots$}
\put(30,55){$\Phi_{n-2}[n-2]$}
\put(60,55){$s[n-2]$}
\put(14,53){\vector(0,-2){8}}
\put(36,53){\vector(-3,-2){14}}
\put(44,53){\vector(3,-2){14}}
\put(66,53){\vector(0,-2){8}}
\put(3,40){$\Phi_n[n-1]$}
\put(30,40){$\Phi_{n-1}[n-1]$}
\put(60,40){$s[n-1]$}
\put(14,38){\vector(1,-1){9}}
\put(40,38){\vector(-3,-2){13}}
\put(40,38){\vector(3,-2){13}}
\put(66,38){\vector(-1,-1){9}}
\put(20,25){$\Phi_{n}[n]$}
\put(52,25){$s[n]$}
\put(26,23){\vector(1,-1){9}}
\put(53,23){\vector(-1,-1){9}}
\put(32,10){$s[n+1]$}
\end{picture}
\caption{Diagram describing how to obtain the $(n+1)$-iterated solution generated by Darboux transformations using one solution $s$ of the SSGE (\ref{SSGE}) and $n+1$ solutions $\Phi_j$ of the LSP (\ref{LSP}) associated with the solution $s$ for the fixed spectral parameters $\lambda_j$, $j=0,1,...,n$.}
\label{Fig1}
\end{figure}
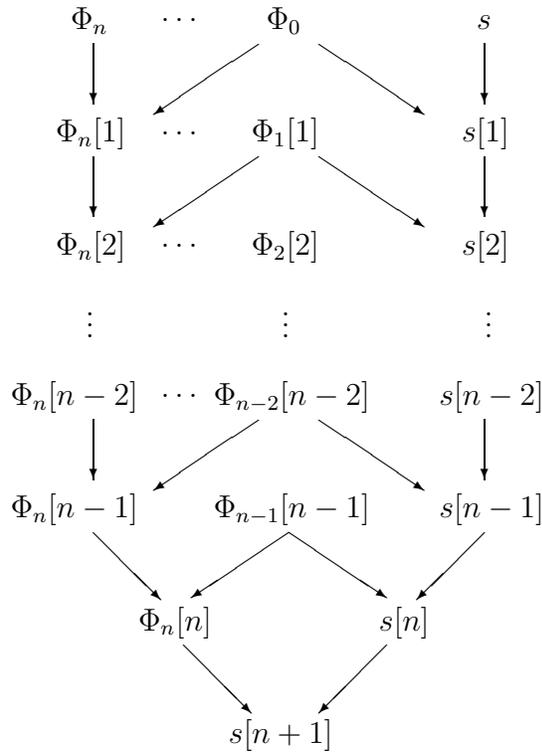

In order to construct a higher iteration solution of the SSGE (\ref{SSGE}), we must ``drop'' other solutions of the LSP associated with $s$ and $\lambda_j$ for the lowest indices $j$. As an example, let us say that we know a solution $s$ of the SSGE and three solutions $\Phi_0$, $\Phi_1$ and $\Phi_2$ of the LSP associated with the fixed-valued spectral parameters $\lambda_0$, $\lambda_1$ and $\lambda_2$, respectively. Hence, from $\Phi_1$ and $\Phi_2$, and dropping $\Phi_0$, we get respectively $\Phi_1[1]$ and $\Phi_2[1]$. To iterate once more, we can drop $\Phi_1[1]$ to obtain $\Phi_2[2]$. Moreover, from $\Phi_0$, $\Phi_1[1]$ and $\Phi_2[2]$, we can construct three new solutions $s[1]$, $s[2]$ and $s[3]$, respectively. The procedure can be applied $n$ times using $n$ fixed solutions of the LSP and the associated solution of the SSGE (\ref{SSGE}) as described in Figure \ref{Fig1}.

The second Darboux Transformation of the a solution $s$ of the SSGE (\ref{SSGE}) is given by
\begin{equation}
s[2]=s-i\ln\left[\frac{\lambda_0\phi_0\psi_1-\lambda_1\phi_1\psi_0+i\sqrt{\lambda_0\lambda_1}\chi_0\chi_1}{\lambda_0\phi_1\psi_0-\lambda_1\phi_0\psi_1+i\sqrt{\lambda_0\lambda_1}\chi_0\chi_1}\right].
\end{equation}

By repeating the Darboux transformation $n$ times ($n>2$), we obtain by induction the solution $s[n]$ which takes the form
\begin{equation}
\hspace{-2.5cm}
\begin{array}{l}
s[n]=s-i\ln\left[\frac{\sum^{(n+1)/2}_{m=0}(i)^mP(\lambda_{k_j};\lambda_{k_\nu})\Delta^1_{k_j}X_{k_\nu}}{\sum^{(n+1)/2}_{m=0}(i)^mP(\lambda_{k_j};\lambda_{k_\nu})\Delta^2_{k_j}X_{k_\nu}}\right],\quad \mbox{when $n$ is odd,}\\
s[n]=s-i\ln\left[\frac{\sum^{n/2}_{m=0}(i)^mP(\lambda_{k_j};\lambda_{k_\nu})\Delta^1_{k_j}X_{k_\nu}+\sum_{m=1}^{n/2-1}P(\lambda_{k_j};\lambda_{k_\nu})\frac{(-1)^{m}}{m!}X_{k_j}X_{k_\nu}}{\sum^{n/2}_{m=0}(i)^mP(\lambda_{k_j};\lambda_{k_\nu})\Delta^2_{k_j}X_{k_\nu}+\sum_{m=1}^{n/2-1}P(\lambda_{k_j};\lambda_{k_\nu})\frac{(-1)^{m}}{m!}X_{k_j}X_{k_\nu}}\right],\quad \mbox{when $n$ is even,}
\end{array}
\label{soln}
\end{equation}
where $k_j$ represents the set of $n-2m$ ordered indices and $k_\nu$ represents the set of $2m$ ordered indices not in $k_j$. The quantities $\Delta^i_{abc...}$, $X_{abc...}$ and $P(\lambda_a,\lambda_b,...;\lambda_c,\lambda_d...)$ are given by
\begin{equation}
\hspace{-1cm}\begin{array}{l}
\Delta^1_{abc...}=\det\left(\begin{array}{cccc}
\vdots & \vdots & \vdots & \\
\lambda_a^2\psi_a & \lambda_b^2\psi_b & \lambda_c^2\psi_c & \cdots \\
\lambda_a\phi_a & \lambda_b\phi_b & \lambda_c\phi_c & \cdots \\
\psi_a & \psi_b & \psi_c & \cdots
\end{array}\right),\\
\Delta^2_{abc...}=\det\left(\begin{array}{cccc}
\vdots & \vdots & \vdots & \\
\lambda_a^2\phi_a & \lambda_b^2\phi_b & \lambda_c^2\phi_c & \cdots \\
\lambda_a\psi_a & \lambda_b\psi_b & \lambda_c\psi_c & \cdots \\
\phi_a & \phi_b & \phi_c & \cdots
\end{array}\right),\\
\\
X_{abcd...}=\sqrt{\lambda_a\lambda_b\lambda_c\lambda_d...}\chi_a\chi_b\chi_c\chi_d...,\\
\\
P(\lambda_a,\lambda_b,...;\lambda_c,\lambda_d...)=(-1)^{n+\alpha}(\lambda_a+\lambda_c)(\lambda_a+\lambda_d)...(\lambda_b+\lambda_c)(\lambda_b+\lambda_d)...,
\end{array}
\end{equation}
where $P(...\,;...)$ has the following definition when it has no argument $\lambda_k$ on one side or the other of the semicolon
\begin{equation}
P(\lambda_{k_j};\emptyset)=1=P(\emptyset;\lambda_{k_\nu}),
\end{equation}
and $\alpha$ is the binary function
\begin{equation}
\alpha=\left\lbrace\begin{array}{cl}
0 & \begin{array}{c}
\mbox{if }ab...cd...\mbox{ are equivalent to an even number}\\
\mbox{ of cyclic permutations of the ordered indices,}
\end{array} \\
\\
1 & \begin{array}{c}
\mbox{if }ab...cd...\mbox{ are equivalent to an odd number}\\
\mbox{ of cyclic permutations of the ordered indices.}
\end{array}
\end{array}\right.
\end{equation}
In addition, we define $\Delta_\emptyset=0$.

As examples, we provide the first four Darboux transformations of a known solution $s$ of the SSGE (\ref{SSGE}):
\begin{equation*}
s[1]=s-i\ln\left(\frac{\Delta_0^1}{\Delta_0^2}\right),
\end{equation*}
\begin{equation*}
s[2]=s-i\ln\left(\frac{\Delta_{01}^1+iX_{01}}{\Delta_{01}^2+iX_{01}}\right),
\end{equation*}
\begin{equation*}
\hspace{-2.5cm}s[3]=s-i\ln\left(\frac{\Delta_{012}^1+iP(\lambda_0;\lambda_1,\lambda_2)\Delta_0^1X_{12}+iP(\lambda_2;\lambda_0,\lambda_1)\Delta_2^1X_{01}+iP(\lambda_1;\lambda_0,\lambda_2)\Delta_1^1X_{02}}{\Delta_{012}^2+iP(\lambda_0;\lambda_1,\lambda_2)\Delta_0^2X_{12}+iP(\lambda_2;\lambda_0,\lambda_1)\Delta_2^2X_{01}+iP(\lambda_1;\lambda_0,\lambda_2)\Delta_1^2X_{02}}\right),
\end{equation*}
\begin{equation*}
\hspace{-2.5cm}\begin{array}{l}
s[4]=s-i\ln\left[\left(\Delta_{0123}^1+iP(\lambda_0,\lambda_3;\lambda_1,\lambda_2)\Delta_{03}^1X_{12}+iP(\lambda_0,\lambda_2;\lambda_1,\lambda_3)\Delta_{02}^1X_{13}\right.\right.\\
+iP(\lambda_0,\lambda_1;\lambda_2,\lambda_3)\Delta_{01}^1X_{23}+iP(\lambda_1,\lambda_3;\lambda_0,\lambda_2)\Delta_{13}^1X_{02}+iP(\lambda_1,\lambda_2;\lambda_0,\lambda_3)\Delta_{12}^1X_{03}\\
+iP(\lambda_2,\lambda_3;\lambda_0,\lambda_1)\Delta_{23}^1X_{01}\left.-(P(\lambda_0,\lambda_1;\lambda_2,\lambda_3)+P(\lambda_0,\lambda_2;\lambda_1,\lambda_3)+P(\lambda_0,\lambda_3;\lambda_1,\lambda_2))X_{0123}\right)\\
\left./\left(\Delta_{0123}^2+iP(\lambda_0,\lambda_3;\lambda_1,\lambda_2)\Delta_{03}^2X_{12}+iP(\lambda_0,\lambda_2;\lambda_1,\lambda_3)\Delta_{02}^2X_{13}\right.\right.\\
+iP(\lambda_0,\lambda_1;\lambda_2,\lambda_3)\Delta_{01}^2X_{23}+iP(\lambda_1,\lambda_3;\lambda_0,\lambda_2)\Delta_{13}^2X_{02}+iP(\lambda_1,\lambda_2;\lambda_0,\lambda_3)\Delta_{12}^2X_{03}\\
+iP(\lambda_2,\lambda_3;\lambda_0,\lambda_1)\Delta_{23}^2X_{01}\left.\left.-(P(\lambda_0,\lambda_1;\lambda_2,\lambda_3)+P(\lambda_0,\lambda_2;\lambda_1,\lambda_3)+P(\lambda_0,\lambda_3;\lambda_1,\lambda_2))X_{0123}\right)\right].
\end{array}
\end{equation*}

\section{Explicit solutions used for the bosonic supersymmetric Sym-Tafel formula for immersion}\label{SecSymTafel}
By considering the trivial solution of the SSGE (\ref{SSGE}),
\begin{equation}
s=2k\pi,\qquad k\in\mathbb{Z}\label{s0}
\end{equation}
the solution $\Phi$ of the LSP for any invertible value of $\lambda_j\in\mathbb{G}$ is given by
\begin{equation}
\hspace{-2cm}\left(\begin{array}{c}
\psi_j\\
\phi_j\\
\chi_j
\end{array}\right)=\left(\begin{array}{c}
c_j+\left(\frac{-b_j}{2\sqrt{\lambda_j}}-\frac{i\underline{a}_j}{2\sqrt{\lambda_j}}\theta^++i\sqrt{\lambda_j}\underline{a}_j\theta^-+\frac{ib_j}{2\sqrt{\lambda_j}}\theta^+\theta^-\right)e^{\eta_j}\\
\\
c_j-\left(\frac{-b_j}{2\sqrt{\lambda_j}}-\frac{i\underline{a}_j}{2\sqrt{\lambda_j}}\theta^++i\sqrt{\lambda_j}\underline{a}_j\theta^-+\frac{ib_j}{2\sqrt{\lambda_j}}\theta^+\theta^-\right)e^{\eta_j}\\
\\
\left(\underline{a}_j+\frac{b_j}{2\lambda_j}\theta^+-b_j\theta^-+i\underline{a}_j\theta^+\theta^-\right)e^{\eta_j}
\end{array}\right),\label{phi0}
\end{equation}
for j=0,1,2,... , where 
\begin{equation}
\eta_j=\frac{x_+}{2\lambda_j}-2\lambda_j x_-
\end{equation} 
is a bosonic linear function of $x_+$ and $x_-$, $\underline{a}_j$ is an arbitrary fermionic constant and $b_j,c_j$ are arbitrary bosonic constants. Since the solution (\ref{phi0}) satisfies the LSP (\ref{LSP}) for any value of $\lambda_j$, it is possible to compute a high number of solutions using the Darboux transformations from equation (\ref{soln}). 

In the further examples, we will only consider two non-trivial solutions using the first iteration of the Darboux transformations for the geometric characterization of the bosonic SUSY version of the Sym-Tafel formula for immersion.

According to \cite{BG16}, we now present the bosonic SUSY Sym-Tafel formula for the immersion of solitonic surfaces in Lie superalgebras.

\begin{proposition}\label{Prop1}
Let us assume that there exists a LSP of the form (\ref{LSP2}) associated with a SUSY integrable systems of partial differential equations $\Omega=0$, where the fermionic potential matrices $U_\pm$ take values in the $\mathfrak{gl}(m\vert n)$ Lie superalgebra and the wavefunction $\Psi$ takes value in the $GL(m\vert n)$ Lie supergroup. Consider the bosonic infinitesimal deformations
\begin{equation}
\tilde{U}_\pm=U_\pm+\epsilon\beta(\lambda)\partial_\lambda U_\pm\in\mathfrak{gl}(m\vert n),\qquad \tilde{\Psi}=\Psi(I+\epsilon F)\in GL(m\vert n),
\end{equation}
that preserve both the LSP (\ref{LSP2}) and the zero-curvature condition (\ref{ZCC}) for an arbitrary bosonic function $\beta(\lambda)$ of $\lambda$, where $\partial_\lambda$ is the derivative with respect to $\lambda$ and $\epsilon$ is a bosonic infinitesimal parameter whose quadratic terms are neglected. Then, there exists an immersion bosonic supermatrix $F$ given by
\begin{equation}
F=\beta(\lambda)\Psi^{-1}\partial_\lambda\Psi\in\mathfrak{gl}(m\vert n)
\end{equation}
which defines a two-dimensional surface in a Lie superalgebra whenever its tangent vectors
\begin{equation}
ED_\pm F=\beta(\lambda)\Psi^{-1}E\partial_\lambda U_\pm\Psi,\qquad E=\left(\begin{array}{cc}
 I_m & 0 \\
0 & -I_n 
\end{array}\right),
\end{equation}
are linearly independent, where $I_m$ and $I_n$ are the identity matrices of dimension $m$ and $n$, respectively.
\end{proposition}

Using the super Killing form defined by the supertrace,
\begin{equation}
\langle A,B\rangle=\frac{1}{2}\mbox{str}(AB)=\frac{1}{2}\mbox{tr}(E^{\deg(AB)+1}AB),
\end{equation}
we obtain that the metric coefficients are given by the relations \cite{BG16}
\begin{equation}
\hspace{-2cm}g_{ii}=\langle \beta(\lambda)E\partial_\lambda U_\pm,\beta(\lambda)\partial_\lambda U_\pm\rangle,\qquad g_{12}=-g_{21}=\langle \beta(\lambda)E\partial_\lambda U_+,\beta(\lambda)E\partial_\lambda U_-\rangle
\end{equation}
where $i=1,2$ and $1,2$ stand for $+,-$ respectively, such that the first fundamental form is given by
\begin{equation}
I=(d_+)^2g_{11}+2d_+d_-g_{12}+(d_-)^2g_{22}.
\end{equation}
According to \cite{BGH151}, the fermionic differential forms $d_\pm$ anticommute with each other,
\begin{equation}
\lbrace d_+,d_-\rbrace=0,
\end{equation}
and represent the infinitesimal displacement in the direction of $D_\pm$. The discriminant $g$ of the metric is given by
\begin{equation}
g=g_{11}g_{22}-g_{12}g_{21}=g_{11}g_{22}+(g_{12})^2.
\end{equation}
The unit normal vector $N$ satisfies the relations
\begin{equation}
\langle N,N\rangle=1,\qquad \langle ED_\pm F,N\rangle=0.
\end{equation}
The vector $N$ can be written only in terms of the tangent vectors,
\begin{equation}
N=\frac{\lbrace ED_+F,ED_-F\rbrace}{\langle\lbrace ED_+F,ED_-F\rbrace,\lbrace ED_+F,ED_-F\rbrace\rangle^{1/2}},
\end{equation}
assuming that the norm,
\begin{equation}
\Vert\lbrace ED_+F,ED_-F\rbrace\Vert=\langle\lbrace ED_+F,ED_-F\rbrace,\lbrace ED_+F,ED_-F\rbrace\rangle^{1/2},
\end{equation}
is invertible. The coefficients of the second fundamental form are given by
\begin{equation}
b_{ij}=\langle D_jD_iF,N\rangle=\langle \beta(\lambda)D_j\partial_\lambda U_i-\lbrace \beta(\lambda)E\partial_\lambda U_i,EU_j\rbrace,\Psi N\Psi^{-1}\rangle,
\end{equation}
where $i,j=1,2$. The indices $1,2$ stand for $+,-$ respectively. The second fundamental form is
\begin{equation}
II=(d_+)^2b_{11}+2d_+d_-b_{12}+(d_-)^2b_{22}.
\end{equation}
The Gaussian and mean curvatures are written respectively as
\begin{equation}
K=\frac{b_{11}b_{22}+(b_{12})^2}{g_{11}g_{22}+(g_{12})^2},\qquad H=\frac{b_{11}g_{22}+b_{22}g_{11}+2b_{12}g_{12}}{2(g_{11}g_{22}+(g_{12})^2)}.
\end{equation}

A first solution $s_1[1]$ of the SSGE (\ref{SSGE}) is constructed using the first Darboux transformation (\ref{sol1}) for the solutions (\ref{s0}) and (\ref{phi0}), where the constant $b_0$ is sent to zero. The solution $s_1[1]$ takes the form
\begin{equation}
s_1[1]=2k\pi-i\ln\left(1-\frac{i\underline{a}_0e^{\eta_0}}{c_0\sqrt{\lambda_0}}\theta^++\frac{2i\sqrt{\lambda_0}}{c_0}\underline{a}_0e^{\eta_0}\theta^-\right).
\end{equation}
The associated solution $\Phi_1$ of the LSP for $\lambda=\lambda_1$ is obtained from the 1-Darboux transformations (\ref{Dar1}),
\begin{equation}
\hspace{-2.6cm}\begin{array}{r}
\psi_1[1]=\left[(\lambda_1-\lambda_0)c_1-\frac{i}{c_0}\sqrt{\lambda_0\lambda_1}\underline{a}_0\underline{a}_1e^{\eta_0+\eta_1}\right]+\left[\frac{i}{2\sqrt{\lambda_1}}(\lambda_1+\lambda_0)\underline{a}_1e^{\eta_1}-i\sqrt{\lambda_0}\frac{c_1}{c_0}\underline{a}_0e^{\eta_0}\right]\theta^+\\
+\left[-i\sqrt{\lambda_1}(\lambda_1+\lambda_0)\underline{a}_1e^{\eta_1}+2i\lambda_0^{3/2}\frac{c_1}{c_0}\underline{a}_0e^{\eta_0}\right]\theta^-+\left[\sqrt{\frac{\lambda_0}{\lambda_1}}(\lambda_1+\lambda_0)\frac{\underline{a}_0\underline{a}_1}{c_0}e^{\eta_0+\eta_1}\right]\theta^+\theta^-,\\
\\
\phi_1[1]=\left[(\lambda_1-\lambda_0)c_1-\frac{i}{c_0}\sqrt{\lambda_0\lambda_1}\underline{a}_0\underline{a}_1e^{\eta_0+\eta_1}\right]+\left[\frac{-i}{2\sqrt{\lambda_1}}(\lambda_1+\lambda_0)\underline{a}_1e^{\eta_1}+i\sqrt{\lambda_0}\frac{c_1}{c_0}\underline{a}_0e^{\eta_0}\right]\theta^+\\
+\left[i\sqrt{\lambda_1}(\lambda_1+\lambda_0)\underline{a}_1e^{\eta_1}-2i\lambda_0^{3/2}\frac{c_1}{c_0}\underline{a}_0e^{\eta_0}\right]\theta^-+\left[\sqrt{\frac{\lambda_0}{\lambda_1}}(\lambda_1+\lambda_0)\frac{\underline{a}_0\underline{a}_1}{c_0}e^{\eta_0+\eta_1}\right]\theta^+\theta^-,\\
\\
\chi_1[1]=\left[-(\lambda_0+\lambda_1)\underline{a}_1e^{\eta_1}+2\sqrt{\lambda_0\lambda_1}\frac{c_1}{c_0}\underline{a}_0e^{\eta_0}\right](1+i\theta^+\theta^-).\hfill
\end{array}
\end{equation}
The associated pseudo-Riemannian geometry taken from the bosonic SUSY version of the Sym-Tafel immersion formula gives the following coefficients for the metric
\begin{equation}
g_{11}=\frac{-i}{2\lambda},\qquad g_{12}=-g_{21}=-i,\qquad g_{22}=2i\lambda,
\end{equation}
where the arbitrary function $\beta(\lambda)$ is taken to be $\beta=2\lambda$. The coefficients $b_{ij}$ of the second fundamental form are given by
\begin{equation}
b_{11}=b_{22}=0,\qquad b_{12}=-b_{21}=\frac{\underline{a}_0}{c_0}e^{\eta_0}\left(\frac{-1}{\sqrt{\lambda_0}}\theta^++2\sqrt{\lambda_0}\theta^-\right).
\end{equation}
The Gaussian curvature $K=1$ implies that the surface can be classified as a constant positive Gaussian curvature one, which would implies that it would be a sphere in comparison with the classical geometry. However, the mean curvature cannot be computed since the discriminant $g=g_{11}g_{22}-g_{12}g_{21}$ vanishes.

A second 1-Darboux transformation solution $s_2[1]$ is considered using the constraint $\underline{a}_0=0$ on solution (\ref{s0}) and (\ref{phi0}), which leads to the particular solution
\begin{equation}
\begin{array}{l}
s_2[1]=2k\pi-i\ln\left[\left(c_0+\frac{b_0}{2\sqrt{\lambda_0}}e^{\eta_0}\right)^{-1}\left(c_0-\frac{b_0}{2\sqrt{\lambda_0}}e^{\eta_0}\right)\right.\\
\hspace{2.5cm}\left.+2c_0\left(c_0+\frac{b_0}{2\sqrt{\lambda_0}}e^{\eta_0}\right)^{-2}\frac{ib_0}{2\sqrt{\lambda_0}}e^{\eta_0}\theta^+\theta^-\right].
\end{array}
\end{equation}
The associated solution $\Phi_2$ of the LSP for $\lambda=\lambda_1$ is given by
\begin{equation}
\hspace{-2.5cm}\begin{array}{l}
\psi_1[1]=\left[\lambda_1c_1+\frac{\sqrt{\lambda_1}b_1}{2}e^{\eta_1}-\lambda_0\left(c_0+\frac{b_0}{2\sqrt{\lambda_0}}e^{\eta_0}\right)\left(c_1-\frac{b_1}{2\sqrt{\lambda_1}}e^{\eta_1}\right)\left(c_0-\frac{b_0}{2\sqrt{\lambda_0}}e^{\eta_0}\right)^{-1}\right]\\
\hspace{1.5cm}+\left[\left(c_1-\frac{b_1}{2\sqrt{\lambda_1}}e^{\eta_1}\right)\left(c_0-\frac{b_0}{2\sqrt{\lambda_0}}e^{\eta_0}\right)^{-1}\frac{i\sqrt{\lambda_0}b_0}{2}e^{\eta_0}\right.\\
\hspace{1.5cm}-\lambda_0\left(c_0+\frac{b_0}{2\sqrt{\lambda_0}}e^{\eta_0}\right)\left(c_0-\frac{b_0}{2\sqrt{\lambda_0}}e^{\eta_0}\right)^{-1}\frac{ib_1}{2\sqrt{\lambda_1}}+\frac{-i\sqrt{\lambda_1}b_1}{2}e^{\eta_1}\\
\hspace{1.5cm}\left.+\left(c_0+\frac{b_0}{2\sqrt{\lambda_0}}e^{\eta_0}\right)\left(c_1-\frac{b_1}{2\sqrt{\lambda_1}}e^{\eta_1}\right)\left(c_0-\frac{b_0}{2\sqrt{\lambda_0}}e^{\eta_0}\right)^{-2}\frac{ib_0\sqrt{\lambda_0}}{2}e^{\eta_0}\right]\theta^+\theta^-,\\
\\
\phi_1[1]=\left[\lambda_1c_1-\frac{\sqrt{\lambda_1}b_1}{2}e^{\eta_1}-\lambda_0\left(c_0-\frac{b_0}{2\sqrt{\lambda_0}}e^{\eta_0}\right)\left(c_1+\frac{b_1}{2\sqrt{\lambda_1}}e^{\eta_1}\right)\left(c_0+\frac{b_0}{2\sqrt{\lambda_0}}e^{\eta_0}\right)^{-1}\right]\\
\hspace{1.5cm}+\left[\lambda_0\left(c_0-\frac{b_0}{2\sqrt{\lambda_0}}e^{\eta_0}\right)\left(c_0-\frac{b_0}{2\sqrt{\lambda_0}}e^{\eta_0}\right)^{-1}\frac{ib_1}{2\sqrt{\lambda_1}}e^{\eta_1}\right.\\
\hspace{1.5cm}-\left(c_1+\frac{b_1}{2\sqrt{\lambda_1}}e^{\eta_1}\right)\left(c_0+\frac{b_0}{2\sqrt{\lambda_0}}e^{\eta_0}\right)^{-1}\frac{i\sqrt{\lambda_0}b_0}{2}e^{\eta_0}+\frac{i\sqrt{\lambda_1}b_1}{2}e^{\eta_1}\\
\hspace{1.5cm}\left.-\left(c_0-\frac{b_0}{2\sqrt{\lambda_0}}e^{\eta_0}\right)\left(c_1+\frac{b_1}{2\sqrt{\lambda_1}}e^{\eta_1}\right)\left(c_0+\frac{b_0}{2\sqrt{\lambda_0}}e^{\eta_0}\right)^{-2}\frac{i\sqrt{\lambda_0}b_0}{2}e^{\eta_0}\right]\theta^+\theta^-,\\
\\
\vspace{2mm}\chi_1[1]=\left[-(\lambda_0+\lambda_1)\frac{b_1}{2\lambda_1}e^{\eta_1}+\sqrt{\frac{\lambda_1}{\lambda_0}}b_0e^{\eta_0}\left(c_0^2-\frac{b_0^2}{4\lambda_0}e^{2\eta_0}\right)^{-1}\left(c_0c_1-\frac{b_0b_1}{4\sqrt{\lambda_0\lambda_1}}e^{\eta_0+\eta_1}\right)\right]\theta^+\\

\hspace{1cm}+\left[(\lambda_0+\lambda_1)b_1e^{\eta_1}-2\sqrt{\lambda_0\lambda_1}b_0e^{\eta_0}\left(c_0^2-\frac{b_0^2}{4\lambda_0}e^{2\eta_0}\right)^{-1}\left(c_0c_1-\frac{b_0b_1}{4\sqrt{\lambda_0\lambda_1}}e^{\eta_0+\eta_1}\right)\right]\theta^-.
\end{array}
\end{equation}
The first fundamental form's coefficients $g_{ij}$ for the associated pseudo-Riemannian geometry of the bosonic SUSY version of the Sym-Tafel immersion formula with $\beta(\lambda)=2\lambda$ are given by
\begin{equation}
\hspace{-1cm}\begin{array}{l}
g_{11}=\frac{-i}{2\lambda}\qquad g_{22}=2i\lambda\\
g_{12}=-g_{21}=-i\cos(s_2[1])\\
\phantom{g_{12}}=-i\left(c_0^2-\frac{b_0^2}{4\lambda_0}e^{2\eta_0}\right)^{-1}\left(c_0^2+\frac{b_0^2}{4\lambda_0}e^{2\eta_0}\right)-2\left(c_0^2-\frac{b_0^2}{4\lambda_0}^{2\eta_0}\right)^{-2}\frac{c_0^2b_0^2}{\lambda_0}e^{\eta_0}\theta^+\theta^-
\end{array}
\end{equation}
and the coefficients $b_{ij}$ of the second fundamental form are
\begin{equation}
\hspace{-1cm}\begin{array}{l}
b_{11}=b_{22}=0,\\
b_{12}=-b_{21}=\sin(s_2[1])\\
\phantom{b_{12}}=i\left(c_0^2-\frac{b_0^2}{4\lambda_0}e^{2\eta_0}\right)^{-1}\frac{c_0b_0}{\sqrt{\lambda_0}}e^{\eta_0}+\left(c_0^2-\frac{b_0^2}{4\lambda_0}e^{2\eta_0}\right)^{-2}\left(c_0^2+\frac{b_0^2}{4\lambda_0}e^{2\eta_0}\right)\frac{c_0b_0}{\sqrt{\lambda_0}}e^{\eta_0}.
\end{array}
\end{equation}
Both discriminants $g=g_{11}g_{22}-g_{12}g_{21}$ and $b=b_{11}b_{22}-b_{12}b_{21}$ are equal to
\begin{equation}
\begin{array}{l}
\sin^2(s_2[1])=-\left(c_0^2-\frac{b_0^2}{4\lambda_0}e^{2\eta_0}\right)^{-2}\frac{c_0^2b_0^2}{\lambda_0}e^{2\eta_0}\\
\hspace{2cm}+2i\left(c_0^2-\frac{b_0^2}{4\lambda_0}e^{2\eta_0}\right)^{-3}\left(c_0^2+\frac{b_0^2}{4\lambda_0}e^{2\eta_0}\right)\frac{c_0^2b_0^2}{\lambda_0}e^{2\eta_0}\theta^+\theta^-.
\end{array}
\end{equation}
The Gaussian curvature $K$ is equal to $1$ and the mean curvature takes the non-trivial form
\begin{equation}
\begin{array}{l}
H=-i\cot(s_2[1]),\\
\phantom{H}=-\frac{\sqrt{\lambda_0}}{c_0b_0}e^{-\eta_0}\left(c_0^2-\frac{b_0^2}{4\lambda_0}e^{2\eta_0}\right)-2\left(c_0^2-\frac{b_0^2}{4\lambda_0}e^{2\eta_0}\right)^{-1}\frac{c_0b_0}{\sqrt{\lambda_0}}e^{\eta_0}\theta^+\theta^-\\
\phantom{H+++}+\left(c_0^2-\frac{b_0^2}{4\lambda_0}e^{2\eta_0}\right)^{-1}\left(c_0^2+\frac{b_0^2}{4\lambda_0}e^{2\eta_0}\right)^2\frac{\sqrt{\lambda_0}}{c_0b_0}e^{-\eta_0}\theta^+\theta^-,
\end{array}
\end{equation}
in terms of the bosonic quantities $\lambda_0,x_+,x_-$ and $\theta^+\theta^-$. By considering the case where $b_0=2\sqrt{\lambda_0}c_0$, we obtain that the body part of the mean curvature is simply given by
\begin{equation}
H_b=\sinh\eta_0.
\end{equation}

\section{Conclusions}\label{Conc}
In this paper, we study the links between some of the integrability properties associated with the SSGE. First, we derive fermionic potential supermatrices in $\mathfrak{sl}(2\vert1,\mathbb{G})$ which provide a LSP whose zero-curvature condition corresponds to the SSGE. Using this LSP, we construct an equivalent LSP in terms of bosonic derivatives instead of fermionic derivatives, which require that the potential matrices be bosonic supermatrices in $\mathfrak{sl}(2\vert1,\mathbb{G})$. Moreover, we provide links between the fermionic LSP, coupled sets of super Riccati equations whose compatibility condition is equivalent to the SSGE, and the associated auto-B\"acklund transformation. Furthermore, we provide a comprehensive description of the Darboux transformation associated with the SSGE. This Darboux transformation allows us to provide non-trivial multisoliton solutions of the SSGE.

The bosonic SUSY version of the Sym-Tafel formula for immersion is investigated through examples for the SSGE. Using 1-Darboux transformation solutions, we are able to compute two new examples of geometric characterizations of the associated surfaces immersed in the Lie superalgebra $\mathfrak{gl}(2\vert1,\mathbb{G})$ exclusively in terms of the fermionic and bosonic independent variables. These two surfaces are linked with spheres in analogy with the classical differential geometry since they have a positive constant Gaussian curvature, $K=1$.

The subjects addressed in this paper can be extended in many directions. Among them, we can study other SUSY integrable systems based on their integrability properties and evaluate some examples of immersed surfaces in Lie superalgebras. Moreover, it would be interesting to find an invertible wavefunction $\Psi$ so that we could explicitly compute the deformed surfaces $F$, which are written in terms of the wavefunction $\Psi$. From these surfaces, it should be possible to graphically show the shape of the surfaces and see how their characteristics, such as the metric and curvatures, manifest themselves.

\section*{Acknowledgements}
SB has been partially supported by a doctoral fellowship provided by the FQRNT of the Gouvernement du Qu\'ebec. SB wishes to thank Professor AM Grundland (Centre de Recherches Math\'ematiques, Universit\'e de Montr\'eal) for his useful discussions and support.

\end{document}